\documentclass[a4paper,11pt]{article}
\pdfoutput=1
\usepackage[utf8]{inputenc}
\usepackage[T1]{fontenc}
\usepackage[english]{babel}

\usepackage{indentfirst}
\usepackage{amsmath}
\usepackage{amsthm}
\usepackage{amssymb}
\usepackage{graphicx}
\usepackage{psfrag}
\usepackage{color}
\usepackage{pgf,pgfnodes}
\usepackage[font=small]{caption}
\usepackage{subcaption}

\allowhyphens

\newcommand{\ordo}{\mathcal{O}}

\newcommand{\ket}[1]{{\left|#1\right\rangle}}
\newcommand{\bra}[1]{{\left\langle #1\right|}}

\newcommand{\skalarszorzat}[2]{{\langle #1 | #2 \rangle}}

\newcommand{\vev}[1]{\left\langle #1 \right\rangle}

\usepackage{ifpdf}

\ifpdf
\usepackage{epstopdf}
\usepackage[pdftex,colorlinks,urlcolor=blue,citecolor=blue,linkcolor=blue]{hyperref}
\else
\usepackage[hypertex,colorlinks,urlcolor=blue,citecolor=blue,linkcolor=blue]{hyperref}
\fi

\makeatother

\begin{document}

\numberwithin{equation}{section}
%\tableofcontents

\title{Failure of the Generalized Eigenstate Thermalization~Hypothesis
  in integrable models with multiple particle species}
\author{Bal\'azs Pozsgay$^1$\\
~\\
 $^{1}$MTA--BME \textquotedbl{}Momentum\textquotedbl{} Statistical
Field Theory Research Group\\
1111 Budapest, Budafoki út 8, Hungary
}
\maketitle

\abstract{It has been recently observed for a particular quantum quench
in the XXZ spin chain that local observables do not equilibrate to the
predictions of the Generalized Gibbs Ensemble (GGE). In this work we argue
that the breakdown of the GGE can be attributed to the failure of the
Generalized Eigenstate Thermalization Hypothesis (GETH), which has
been the main candidate to explain the validity of the GGE. We provide
explicit counterexamples to the GETH and argue that generally it does 
not hold in models with multiple particle species. Therefore there is no reason
to assume that the GGE should describe the long time limit of
observables in these integrable models.
}

\section{Introduction}

The problem of equilibration and thermalization of isolated quantum systems has
received considerable interest over the last couple of years
\cite{Silva-quench-colloquium,huse-review}. One of 
the main questions is how the principles of statistical physics can
be derived from the unitary time evolution of the quantum mechanical
model. Interest in these questions has been sparked by new
experimental techniques (for example with cold atoms \cite{cold-atom-review}) where the systems
are almost perfectly isolated from the environment and therefore
 equilibration induced by the system itself can be studied. 

Equilibration in a quantum mechanical system means that the
expectation values of physical observables are convergent as a function of
time in the long time limit. 
Thermalization happens when these stationary values coincide with
those obtained from a thermal ensemble. The full system never thermalises as the unitary time
evolution conserves all information about the initial state. On the
other hand, the reduced density matrices of subsystems can 
approach their thermal values. Physically this means that the full
system can act as a thermal bath for its subsystems and therefore
expectation values of local observables indeed approach thermal
predictions in the long time limit.

One of the main candidates to explain why thermalization can happen is
the Eigenstate Thermalization Hypothesis (ETH)
\cite{eth1,eth2,rigol-eth}. It states that in a typical 
interacting quantum system the excited states which are close in energy
 have approximately the same local correlation functions. 
Supplied with the assumption that in the long-time limit dephasing
between the eigenstates occurs the ETH assures
that all local observables will approach thermal mean values with an
effective temperature determined by the mean value of the energy in
the initial state.

The situation is different in one-dimensional integrable models which possess a family
of higher conserved charges which prevent thermalization in the usual
sense. It was proposed in \cite{rigol-gge} that correlation
functions in integrable models approach values predicted by a Generalized
Gibbs Ensemble (GGE) which incorporates all higher charges with
appropriately chosen Lagrange-multipliers. Furthermore it was proposed
in \cite{rigol-GETH} that equilibration to the GGE
can be explained by a Generalized Eigenstate Thermalization Hypothesis
(GETH). According to the GETH the eigenstates which share the same set of
conserved charges give approximately the same correlation
functions. 
Both the GETH and the GGE are expected to 
become exact in the thermodynamic limit. 

Since its inception the idea of the GGE has become widely excepted in
the field, partly because it was proven to be valid for
models equivalent to free fermions
\cite{free-gge-1,free-gge-2,free-gge-3,free-gge-4,ising-quench-1,ising-quench-2,essler-truncated-gge}. 
However, it was found in \cite{JS-oTBA} that for a certain quantum
quench in the interacting spin-1/2 XXZ chain the GGE gives different predictions
than the quench action (QA) method \cite{quench-action} which (as opposed to
the GGE) is built on first principles and is not based on any
assumptions or approximations. Furthermore clear evidence was found
in a case of a different quench problem in
\cite{sajat-oTBA} that while the predictions of the QA method coincide with
results of real-time simulations, the GGE predictions are not correct.
Differences between the GGE
and real-time simulations were already observed in
\cite{fagotti-collura-essler-calabrese}, but at that time they
were interpreted as the result of very long relaxation times which were
beyond the reach of the simulations. However, the fact that the QA
method (which is exact in the thermodynamic limit) correctly describes
all local correlators \cite{sajat-oTBA} shows that it is the GGE itself
which fails for these particular quench problems. 

In \cite{andrei-gge} it was argued that the breakdown of the GGE
can be explained by the fact that in Bethe Ansatz solvable models with bound
states the set of higher conserved charges is not enough to determine
the distribution of the pseudo-momenta of the particles. Therefore,
states with different root distributions can share the same set of
conserved charges, and still have different correlation
functions. Examples for this had already been observed in
\cite{JS-oTBA}, where it was shown that the QA method
selects states which have the same conserved charges as the initial state,
and still their correlation functions differ from the GGE predictions,
even though the GGE states share the same set of charges by
definition. 

In the present work we take the argument of \cite{andrei-gge} further
and show that the failure of the GGE can be attributed to 
 the failure of the Generalized Eigenstate Hypothesis in models with
 multiple particle species. In the case of the XXZ spin chain we 
 develop a method to construct an arbitrary large family of
 eigenstates which share the same set of local conserved charges and
 still give different local correlations. We give explicit numerical
 examples for the failure of the GETH in the thermodynamic limit. It
 follows from our results that in a typical case the GGE can not
 be a valid description of the long time limit behaviour of the
 system, irrespective of the initial state.

The article is organized as follows. In Section \ref{sec:elso} we
introduce the concepts of thermalization, ETH, GGE and GETH. In
Section \ref{sec:xxzelso} we show that the GETH is not valid in the XXZ
chain, and argue that this is a generic property of models with multiple particle species. In
Section \ref{sec:xxz} we provide explicit counterexamples of the GETH which
are related to a certain quantum quench problem. Section
\ref{sec:conclusions} includes our conclusions, and technical details
about the calculation of correlation functions and the solution of the
so-called TBA system with thermal asymptotics are presented in Appendices
\ref{sec:appcorr} and \ref{sec:iter}, respectively. 

\section{Thermalization in non-integrable and integrable models}

\label{sec:elso}

Consider a generic local Hamiltonian $H$ in a finite volume $L$ with periodic
boundary conditions. To be specific here we treat finite lattice models,
but most of the arguments carry over to continuous models and field
theories as well.

Consider the situation where at $t=0$ the system is prepared in the
initial state $\ket{\Psi_0}$ and for $t>0$ it is evolved
unitarily with a local Hamiltonian $H$. Time dependent expectation
values of local observables are then computed as
\begin{equation*}
  \vev{\ordo(t)}=\sum_{n,m}   c_n c_m^* \bra{m}\ordo\ket{n}
  e^{-i(E_n-E_m)t},  
\end{equation*}
where $c_n=\skalarszorzat{n}{\Psi_0}$. In the large time limit,
neglecting degeneracies we obtain the prediction of the so-called
Diagonal Ensemble, where each state is weighted by the squared norm of
its overlap with the initial state:
\begin{equation}
\label{DE}
\lim_{t\to\infty}  \vev{\ordo(t)}=\sum_{n}  |c_n|^2 \bra{n}\ordo\ket{n}.
\end{equation}
In a finite system the limit in the l.h.s. above does not exist and
time averaging is required to obtain the Diagonal Ensemble on the r.h.s.

If the system thermalized then in a large volume all expectation values should be close
to the canonical prediction 
\begin{equation}
\label{micro}
  \vev{\ordo}_T=\frac{\sum_{n} e^{- E_n/T}  \bra{n}\ordo\ket{n}}{\sum_{n} e^{- E_n/T} }
\end{equation}
with a temperature $T$ that is fixed by the requirement
\begin{equation*}
  \vev{H}_T=\bra{\Psi_0}H\ket{\Psi_0}.
\end{equation*}
It is expected that \eqref{DE} and \eqref{micro} become equal in the
thermodynamic limit.

The expressions \eqref{DE} and \eqref{micro} are seemingly unrelated as the coefficients $|c_n|^2$
are typically random and do not coincide with the Boltzmann
weights. However, it can be shown
that in a
large volume $L$ only those states have non-negligible overlap which share the
energy density of the initial state \cite{rigol-eth}: 
\begin{equation}
\label{e1}
  \frac{E_n}{L}\approx \frac{\bra{\Psi_0}H\ket{\Psi_0}}{L},
\end{equation}
and that the width of the distribution of the energy density goes to
zero in the thermodynamic limit at least as fast as
\begin{equation}
\label{width}
  \Delta \left(\frac{E}{L}\right)
=\frac{1}{L}\sqrt{\bra{\Psi_0}H^2\ket{\Psi_0}-(\bra{\Psi_0}H\ket{\Psi_0})^2}\sim \frac{1}{\sqrt{L}}.
\end{equation}
Equation \eqref{width} holds for local Hamiltonians and initial states
$\ket{\Psi_0}$ which satisfy the cluster decomposition
principle. Physically relevant states belong to this class.

The Eigenstate Thermalization Hypothesis (ETH)
\cite{eth1,eth2,rigol-eth} states that all eigenstates 
that are close in energy have almost the same expectation values of
physical observables and therefore 
\begin{equation}
\label{DE2}
\sum_{n}  |c_n|^2 \bra{n}\ordo\ket{n}\approx 
\left(\sum_{n}  |c_n|^2 \right) \bra{n_1}\ordo\ket{n_1}=
\bra{n_1}\ordo\ket{n_1},
\end{equation}
where it is enough to select one sample state $n_1$ which fulfills the
condition \eqref{e1} and $c_1\ne 0$. 

Applying the ETH to the ensemble average
\eqref{micro} we obtain that
\begin{equation}
\label{micro2}
  \vev{\ordo}_T\approx\bra{n_1}\ordo\ket{n_1},
\end{equation}
where we used that in a large volume the canonical ensemble also
selects states which are close in energy, such that their
energy density coincides with that of the initial state. Comparing \eqref{DE2} and
\eqref{micro2} it follows that local observables indeed thermalize:
\begin{equation}
\label{eth}
\lim_{t\to\infty}  \vev{\ordo(t)}\approx
\vev{\ordo}_T.
\end{equation}
An exact equality is expected in the thermodynamic limit.

\subsection{The GGE in integrable models}

If the system is integrable then there exists a family of higher
charges $\{Q_j\}$ such that each member is a sum of local operators
and they all commute and the Hamiltonian is a member of the series. As
a result the expectation values of the $Q_j$ are integrals of motion
which preclude thermalization in the usual sense. Nevertheless, even
integrable models are expected to equilibrate and the question arises
whether some kind of statistical physical ensemble describes the
stationary values.

It was proposed in \cite{rigol-gge} that a Generalized Gibbs Ensemble,
which is the natural extension of the canonical ensemble, should
describe the local observables. To be precise, the
following should hold:
\begin{equation}
\label{GGE}
\lim_{t\to\infty}  \vev{\ordo(t)}=\vev{\ordo \rho_{\text{GGE}}},\text{
  where   }
\rho_{\text{GGE}}=\frac{e^{-\sum_j \lambda_j Q_j}}{\text{Tr}\ e^{-\sum_j \lambda_j Q_j}},
\end{equation}
where the parameters $\lambda_j$ are fixed by the requirement
\begin{equation*}
  \vev{Q_j\rho_{\text{GGE} }}=\bra{\Psi_0}Q_j\ket{\Psi_0},\qquad
  j=1\dots N_Q,
\end{equation*}
where $N_Q$ is the number of the higher charges. In a finite volume we
have typically $N_Q=L$. 

Evidently it is necessary to add all existing local charges to
the GGE. Also, it follows from the usual statistical physical
arguments that only extensive operators can be added to the exponent,
otherwise the thermodynamic limit could not be defined.

It was argued in \cite{rigol-GETH} that a possible mechanism for
thermalization to the GGE is the appropriate extension of the ETH to
the integrable case: the Generalized Eigenstate Thermalization
Hypothesis (GETH). This hypothesis states that if all local conserved charges of two
different eigenstates are close to each other, then the mean values of
all local operators are close too. In other words, the set of the 
conserved charges uniquely determines the correlations in the
state, at least in the thermodynamic limit. For the diagonal ensemble this means that
\begin{equation}
\label{DE3}
\lim_{t\to\infty}  \vev{\ordo(t)}=\sum_{n}  |c_n|^2 \bra{n}\ordo\ket{n}\approx 
\left(\sum_{n}  |c_n|^2 \right) \bra{n_1}\ordo\ket{n_1}=
\bra{n_1}\ordo\ket{n_1},
\end{equation}
where we selected a sample state $n_1$ which fulfills the
conditions $c_1\ne 0$ and  
\begin{equation} 
\label{cond} 
  \frac{\bra{n_1}Q_j\ket{n_1}}{L}\approx
  \frac{\bra{\Psi_0}Q_j\ket{\Psi_0}}{L},\qquad
j=1\dots N_Q.
\end{equation}
In \eqref{DE3} we assumed that only those states have a non-negligible
overlap which fulfil the condition \eqref{cond}. This follows from the
fact the mean values of the charges are conserved in time, and the
width of the distribution of the charge densities goes to zero
according to 
\begin{equation}
\label{width2}
  \Delta \left(\frac{Q_j}{L}\right)
=\frac{1}{L}\sqrt{\bra{\Psi_0}Q_j^2\ket{\Psi_0}-(\bra{\Psi_0}Q_j\ket{\Psi_0})^2}\sim
\frac{1}{\sqrt{L}}.
\end{equation}
Once again we used that the $Q_j$ are sums of local operators and that
$\ket{\Psi_0}$ satisfies the cluster decomposition principle.

By definition, the density matrix $\rho_{\text{GGE}}$ generates states which have the
prescribed charge densities, therefore applying the GETH again we
obtain
\begin{equation*}
\vev{\ordo \rho_{\text{GGE}}} \approx  \bra{n_1}\ordo\ket{n_1},
\end{equation*}
and finally
\begin{equation}
\label{geth}
\lim_{t\to\infty}  \vev{\ordo(t)}\approx
\vev{\ordo  \rho_{\text{GGE}}}.
\end{equation}

The GETH has been checked for a lattice model of hard-core bosons in
\cite{rigol-GETH}, but until recently it was an open question whether it holds in 
 other integrable models. 
In this work we argue that in models with
 multiple particle species the GETH does not hold, and therefore in these
 cases there is
 no reason to assume that the system equilibrates to the GGE
 predictions.

\section{The GETH in the XXZ spin chain}

\label{sec:xxzelso}

The most familiar interacting integrable models are Bethe Ansatz
solvable theories \cite{korepin-book}. A generic feature of these
models is that the scattering processes are elastic and the
multi-particle scattering processes factorize, i.e. they are products
of two-particle scattering events. As a result multi-particle states
are constructed using interacting single-particle wave functions, and
interaction occurs only when two particles exchange positions
\cite{XXX}. In such models individual particles can be characterized
by their particle type and their pseudo-momenta (rapidities). In
the infinite volume limit and finite particle density it is possible
to work with the rapidity distribution functions. It is a generic
property of these models that in the thermodynamic limit the mean values of the local
charges and also the correlation functions can be expressed using the
rapidity distributions alone \cite{korepin-book,LM-sajat,sajat-corr,JS-oTBA}.

Here we argue, following \cite{andrei-gge}, that in a theory with
multiple particle types (be it fundamental particles or bound states
thereof) the GETH does not hold. The main reason for the failure of
the GETH is that the countably infinite number of constraints posed by
the local charges is not enough to fix the rapidity distributions if
there is more than one particle present in the spectrum. As a result,
two different configurations can share the same charges, but yield
different correlation functions.

As an example we consider the spin-1/2 XXZ chain defined by the Hamiltonian
\begin{equation}
  \label{H}
  H_{XXZ}=\sum_{j=1}^{L} \left\{
\sigma_j^x\sigma_{j+1}^{x}+\sigma_j^y\sigma_{j+1}^{y}+\Delta
(\sigma_j^z\sigma_{j+1}^{z}-1)
\right\}.
\end{equation}
We constrain ourselves to the regime $\Delta>1$.

This model can be solved by the different forms of the Bethe Ansatz
\cite{korepin-book}. Single particle states are spin waves over the
ferromagnetic reference state
$\ket{F_+}=\ket{++\dots+}$. Multi-particle states can be formed by
taking into account the factorized scattering between individual spin
waves. The explicit wave function can be written as
\begin{equation}
  \label{BA}
\Psi_{N}(\lambda_1,\dots,\lambda_N|s_1,\dots,s_N)=\sum_{P\in\sigma_N} 
\prod_j 
\left(\frac{\sin(\lambda_{P_j}+i\eta/2)}{\sin(\lambda_{P_j}-i\eta/2)}\right)^{s_j}
\prod_{j>k} \frac{\sin(\lambda_{P_j}-\lambda_{P_k}-i\eta)}{\sin(\lambda_{P_j}-\lambda_{P_k})}.
\end{equation}
Here the variables $\lambda_j$ are the rapidities of the single spin
waves, $s_j$ denote the positions of the down spins, and
it is assumed that $s_j<s_k$ for $j<k$. The parameter $\eta$ is
defined by $\Delta=\cosh(\eta)$.
In a periodic system the rapidities are subject to the Bethe equations
\begin{equation}
  \label{BAe}
\left(\frac{\sin(\lambda_j+i\eta/2)}{\sin(\lambda_j-i\eta/2)}\right)^L
\prod_{k\ne j}
\frac{\sin(\lambda_j-\lambda_k-i\eta)}{\sin(\lambda_j-\lambda_k+i\eta)}=1.
\end{equation}

The canonical set of commuting local charges of the theory
$\{Q_j\}_{j=1\dots L}$ can be constructed using 
the Algebraic Bethe Ansatz \cite{korepin-book}. Their eigenvalues on a
multi-particle state are given by
\begin{equation}
  Q_{j} \ket{\{\lambda\}_N}=\left( \sum_{k=1}^N q_j(\lambda_k)\right)   \ket{\{\lambda\}_N},
\end{equation}
where
\begin{equation}
q_j(u)=-i \left(\frac{\partial}{\partial u}\right)^j 
\log\left(\frac{\sin(\lambda_j+i\eta/2)}{\sin(\lambda_j-i\eta/2)}\right).
\end{equation}
The Hamiltonian itself is the first member of the series. To be more
precise, in the present normalizations we have
\begin{equation}
  H=2\sinh(\eta) Q_1
\end{equation}

If $\Delta\ge 1$ then spin waves can form arbitrary large bound
states which are called ``strings''
\cite{takahashi-book}. For a bound
state of $n$ fundamental spin waves the rapidities are arranged as
 \begin{equation*}
    \{\lambda\}_n=x-\frac{n-1}{2}i\eta+i\delta_1,x-\frac{n-3}{2}i\eta+i\delta_2,
\dots, x+\frac{n-1}{2}i\eta+i\delta_n.
  \end{equation*}
The variable $x\in [-\pi/2,\pi/2]$ is the string center and the 
$\delta_j$ are string deviations which become exponentially small in
the large volume limit. In the present context is useful to regard the
different strings as different particle types.

In the thermodynamic limit it is convenient to introduce densities for
the string centers such that in a volume $L$ the number of $k$-strings
with centers between $\lambda$ and $\lambda+d\lambda$ is
$L\rho_{\text{r},k}(\lambda)/2\pi$.
The magnetization of the system is then given by
\begin{equation*}
 \vev{S^z}=1/2-
\sum_{k=1}^{\infty} k \int \frac{d\lambda}{2\pi} \rho_{\text{r},k}(\lambda).
\end{equation*}
It is also useful to introduce the densities
$\rho_{\text{h},k}(\lambda)$ for the holes,
 which are the generalization of empty levels of a free theory to the
 Bethe Ansatz solvable case. It can be derived from the Bethe
 equations that these functions 
 satisfy the linear equations \cite{takahashi-book}
\begin{equation}
\label{rhoelso}
\rho_{\text{r},k}+\rho_{\text{h},k}= \delta_{k,1}d+
d\star \left(
\rho_{\text{h},k-1}+\rho_{\text{h},k+1}
\right),
\end{equation}
where
\begin{equation}
\label{du}
d(u)=1+2\sum_{n=1}^\infty \frac{\cos(2n u)}{\cosh(\eta n)}
\end{equation}
and the convolution of two functions is defined as
\begin{equation}
  (f\star g)(u)=
\int_{-\pi/2}^{\pi/2} \frac{d\omega}{2\pi} f(u-\omega) g(\omega).
\end{equation}

Instead of working with the individual conserved charges it is useful
to define the generating function \cite{essler-xxz-gge}
\begin{equation}
\label{G}
  G(\lambda)=\sum_{k=1}^\infty \frac{\lambda^{k-1}}{(k-1)!}\vev{Q_k}.
\end{equation}
It was shown in \cite{JS-oTBA} that $G(\lambda)$ can be expressed
using $\rho_{\text{h},1}(\lambda)$ alone. In our conventions the following
equation holds:
\begin{equation}
\label{rho1hG}
  d\star(s_1+\rho_{\text{h},1})=G,
\end{equation}
where
\begin{equation}
  s_1=-\frac{\sinh(\eta)}{\sin(\lambda+i\eta/2)\sin(\lambda-i\eta/2)}.
\end{equation}
Equation \eqref{rho1hG} means that different configurations 
which have the same $\rho_{\text{h},1}$ also share the same set of local
conserved charges. 

Local correlation functions for arbitrary string distributions can be
computed using the method of \cite{sajat-corr}, which uses results
from the theory of factorization of correlation functions in the finite
temperature case
\cite{XXZ-finite-T-factorization,XXZ-factorization-recent-osszefoglalo}. 
\footnote{Although at present the main results of \cite{sajat-corr}
  are only conjectures, they have
been verified numerically in the finite temperature case
\cite{sajat-corr} and also for a non-trivial quench problem in \cite{sajat-oTBA}.}

In the case of generic string distributions we propose to regard the
hole densities $\{\rho_{\text{h},k}\}_{k=1\dots\infty}$ as the fundamental variables describing the
states. This has the advantage that the root densities are easily
obtained from \eqref{rhoelso} by a simple convolution without a need
to solve any linear equation. 
Moreover, all local correlators can be computed from the hole
densities  using the following steps:
\begin{itemize}
\item As a first step compute the root densities
  $\{\rho_{\text{r},k}\}_{k=1\dots\infty}$ from \eqref{rhoelso}. For
  each string type calculate the functions
  \begin{equation*}
    \eta_k(\lambda)=\frac{\rho_{\text{h},k}(\lambda)}{\rho_{\text{r},k}(\lambda)}.
  \end{equation*}
\item Use the functions $\eta_k(\lambda)$ as an input to certain linear
  equations for two series of auxiliary functions $\rho^{(a)}_k$ and
  $\sigma^{(a)}_k$, where $a=0\dots\infty$ and $k$ is the string
  index.
\item Compute the local correlators using certain integrals over the auxiliary functions.
\end{itemize}
The details of this procedure can be found in Appendix
\ref{sec:appcorr} which includes new and more efficient formulae as
compared to those of \cite{sajat-corr}.

We stress that while the local charges only depend on
$\rho_{\text{h},1}$, the correlations depend on all
$\rho_{\text{h},k}$. On a technical level this is why the GETH fails:
fixing $\rho_{\text{h},1}$ leaves the functions $\rho_{\text{h},k}$
with $k>2$ arbitrary and therefore the local correlators are not
specified by the charges only. 

To conclude this section we point out a simple relation between the
hole densities and the overall magnetization. Multiplying the $k$-th
equation of \eqref{rhoelso} with $k$, integrating over the rapidity
and summing over $k$ we obtain 
\begin{equation*}
\vev{S_z}=\frac{1}{2}-\lim_{k\to\infty}\left( k \sum_{j=1}^k\int \frac{\rho_{\text{r},j}}{2\pi}\right)=
 \frac{1}{2}\lim_{k\to\infty}\left( (k-1) \int \frac{\rho_{\text{h},k}}{2\pi} - k
   \int \frac{\rho_{\text{h},k+1}}{2\pi}\right).
\end{equation*}
Therefore the total magnetization is zero for arbitrary
hole densities as long as the limit on the r.h.s. above vanishes.

\section{Examples for the failure of the GETH}

\label{sec:xxz}

In this section we treat a specific quantum quench problem and
explicitly demonstrate the failure of the GETH in this case. The
problem we consider is the quench from the Majumdar-Ghosh dimer
state, i.e. we assume that at $t=0$ the system is prepared in the
translationally invariant combination
\begin{equation}
\label{dimerstate}
  \ket{\Psi(t=0)}=\frac{1}{\sqrt{2}}(1+T)
\left[\otimes_{1}^{L/2} \frac{\ket{+-}-\ket{-+}}{\sqrt{2}}\right],
\end{equation}
where $T$ is the translation operator by one site. This  vector is
one of the ground states of the Majumdar-Ghosh Hamiltonian \cite{MG}.

This particular quench problem has already been considered in the
works \cite{fagotti-collura-essler-calabrese} and \cite{sajat-oTBA},
both of which included numerical results from real-time
simulations. In \cite{fagotti-collura-essler-calabrese} the GGE
predictions for local observables were calculated using the Quantum
Transfer Matrix method (see also
\cite{sajat-xxz-gge,essler-xxz-gge}). Although discrepancies were
found between the GGE and the real-time simulations, they were
interpreted as a result of long relaxation times\footnote{The paper
\cite{fagotti-collura-essler-calabrese}   considered a single dimer
  product state and not the translational invariant combination. In
  this case it is a separate question whether translational invariance
is restored in the long time limit. However, the GGE was supposed to
describe the averaged correlators, whereas the numerical evidence in
both \cite{fagotti-collura-essler-calabrese}  and \cite{sajat-oTBA}
shows that it fails to do so.}. On the other hand,
the exact predictions from the Quench Action (QA) method were 
computed in \cite{sajat-oTBA} and they were found to agree
with the real-time data. Therefore it was concluded in
\cite{sajat-oTBA} that it is the GGE which is not correct in this case.

In subsection \ref{sec:QA} we recall the QA solution of this quench
problem. In \ref{sec:GGE-TBA} we provide the GGE predictions using
the Thermodynamic Bethe Ansatz (TBA) method and show that they agree
with the numerical results of
\cite{fagotti-collura-essler-calabrese}. Finally in \ref{sec:further}
we develop a method to generate root configurations which have the
same conserved charges but different correlators and thus demonstrate
the failure of the GETH.

\subsection{The Quench Action solution}

The Quench Action method developed in \cite{quench-action} is an exact
method (in the thermodynamic limit) which 
selects the eigenstates that dominate the dynamics
of the system in the long time limit. It does so by minimizing the
so-called Quench Action, which is the combination of the exact
overlaps and the micro-canonical entropy associated to each
state. Specified to the case of the XXZ chain, the main steps of the
solution are as follows.

In a large volume the summation over the states in the Diagonal
Ensemble \eqref{DE} can be replaced by a 
functional integral over the densities. In accordance with the proposal
in the previous section we regard the hole densities as the basic
variables and write
\begin{equation}
\label{funcint}
  \lim_{t\to\infty} \vev{\ordo(t)}=
\int \left[\prod_{j=1}^\infty
\mathcal{D}(\rho_{\text{h},j}(\lambda))\right] 
 \bra{\{\rho_{\text{h}}\}}\ordo \ket{\{\rho_{\text{h}}\}} e^{-L S[\{\rho_{\text{h},j}\}]},
\end{equation}
where $S[\{\rho_{\text{h},j}\}]$ is the Quench Action.
If the exact finite volume overlaps can be written as
\begin{equation*}
  |\skalarszorzat{\Psi_0}{\{\lambda\}_N}|^2=C \prod_{j=1}^N v(\lambda_j),
\end{equation*}
where $C=\ordo(L^0)$, then the QA is expressed as
\begin{equation}
\label{QA}
  S[\{\rho_{\text{h},j}\}]=-\sum_{j=1}^\infty \int_{-\pi/2}^{\pi/2} \frac{du}{2\pi} \left(
\rho_{\text{r},j}(\lambda) g_j(\lambda)+\frac{1}{2} s_j(\lambda)
\right),
\end{equation}
where the $g_j$ and $s_j$ are the overlap and entropy terms for
the $j$-strings:
\begin{equation*}
\begin{split}
  g_j(\lambda)&=\sum_{k=1}^j
  \log\big(v(\lambda+i\eta(n+1-2k)/2)\big)\\
s_j(\lambda)&=\rho_{\text{r},j}(\lambda)\log\frac{\rho_{\text{r},j}(\lambda)+\rho_{\text{h},j}(\lambda)}{\rho_{\text{r},j}(\lambda)}+\rho_{\text{h},j}(\lambda)\log\frac{\rho_{\text{r},j}(\lambda)+\rho_{\text{h},j}(\lambda)}{\rho_{\text{h},j}(\lambda)},
\end{split}
\end{equation*}
and it is understood that the root densities are calculated from the
hole densities using \eqref{rhoelso}.
The factor of $1/2$ in front of the entropy in \eqref{QA} takes into
account that the exact overlaps are non-zero only if the state is
exactly parity symmetric, i.e. it consists of rapidity pairs $\{\pm
\lambda\}$ \cite{Caux-Neel-overlap2,caux-stb-LL-BEC-quench}.  Exact
overlaps were calculated for the N\'eel initial state in
\cite{Caux-Neel-overlap1,Caux-Neel-overlap2}, whereas for the dimer
state it was obtained in \cite{sajat-oTBA} using the results of \cite{Caux-Neel-overlap1,sajat-neel}
that
\begin{equation*}
  v(\lambda)=\frac{\sinh^4(\eta/2)\cot^2(\lambda)}{\sin(\lambda+2i\eta)\sin(\lambda-2i\eta)}.
\end{equation*}

The expression \eqref{funcint} is evaluated in the saddle point
approximation which is exact in the thermodynamic limit. Local
operators do not shift the position of the saddle point, which is therefore 
obtained by the minimalization of the Quench Action itself.
The appropriate generalization of the Thermodynamic Bethe Ansatz (TBA)
 method \cite{takahashi-book}  leads to the following set of
equations for $\eta_j=\rho_{\text{h},j}/\rho_{\text{r},j}$  \cite{JS-oTBA}:
\begin{equation}
\label{oTBA}
  \log\eta_j=f_j
 +d\star \left[
\log(1+\eta_{j-1})+\log(1+\eta_{j+1})
\right],
\end{equation}
where
\begin{equation}
\label{fj}
  f_j=-g_j+d\star(g_{j-1}+g_{j+1}), \text{ with } g_0=0.
\end{equation}
Having found the solution of \eqref{oTBA} the string densities
are calculated from \eqref{rhoelso} and correlators can be
computed using the method described in \cite{sajat-corr} and Appendix
\ref{sec:appcorr}. This task was performed
in
\cite{sajat-oTBA} for different values of $\Delta$ and the QA
predictions were found to be in perfect agreement with 
results of real-time simulations
\footnote{A different form of the overlap-TBA equations  was
used in \cite{sajat-oTBA} where all equations are coupled and a
Lagrange-multiplier fixing the overall magnetization is
present. However, the two formulations lead to the same numerical results
\cite{JS-private}. }. Also, it was found that the saddle point
solution $\{\rho^{QA}_{\text{h},j}\}$
yields the correct local charges and that
\begin{equation}
\label{SumRule}
   S[\{\rho^{QA}_{\text{h},j}\}]=0.
\end{equation}
The vanishing of the Quench Action at the saddle point solution is a
very strong physical requirement which follows from the normalization
of the initial state \cite{sajat-oTBA}:
\begin{equation}
\label{norma}
1= \skalarszorzat{\Psi_0}{\Psi_0}=
\int \left[\prod_{j=1}^\infty
\mathcal{D}(\rho_{\text{h},j}(\lambda))\right] 
e^{-L S[\{\rho_{\text{h},j}\}]}.
\end{equation}
It follows from \eqref{norma} that any root distribution which yields
a positive QA has zero spectral weight in the $L\to\infty$ limit and
therefore it is irrelevant for the dynamics of the system.

\label{sec:QA}

\subsection{The GGE-TBA predictions}

\label{sec:GGE-TBA}

Here we provide the TBA solution of the GGE for the quantum quench
from the dimer state. 
The goal is to compute the local observables from the GGE as
\begin{equation}
\vev{\ordo}_{\text{GGE}} =
\frac{\text{Tr}\left( \ordo e^{-\sum_j \lambda_j Q_j}\right)}
{\text{Tr}\left(e^{-\sum_j \lambda_j Q_j}\right)}.
\end{equation}
The standard treatment of the TBA method selects states with particle
distributions $\rho_{\text{r},k}$,  $\rho_{\text{h},k}$ which minimize
the generalized free energy obtained from the generalized Boltzmann
weights. The following equation is obtained for the functions
$\eta_k$ \cite{JS-oTBA}:
\begin{equation}
\label{GGE-TBA-0}
  \log\eta_j=\delta_{j,1} \left(
-\sum_{k=1}^\infty \lambda_k d^{(k-1)}
\right) +d\star \left[
\log(1+\eta_{j-1})+\log(1+\eta_{j+1})
\right],
\end{equation}
where
\begin{equation*}
  d^{(a)}(u)=\left(\frac{\partial }{\partial u}\right)^{a} d(u)
\end{equation*}
and $d(u)$ is given by \eqref{du} and it is understood that $\eta_0=0$. 
The Lagrange multipliers should be 
fixed by the requirement that the solution of \eqref{GGE-TBA} together
with the equation for the densities \eqref{rhoelso} yields the correct
local charges. It follows from \eqref{rho1hG} that it is enough to fix
the function $\rho_{\text{h},1}$ and this can be used to solve the
GGE-TBA without the parameters $\lambda_j$ \cite{JS-oTBA}. The details
of this method were not given in \cite{JS-oTBA}, therefore we describe
the method that we used.

The two  system of equations
\eqref{GGE-TBA-0} and \eqref{rhoelso} can be considered simultaneously. From the first
equation in \eqref{rhoelso} we obtain 
\begin{equation}
    \rho_{\text{h},1}(1+1/\eta_1)=d+d\star \frac{\rho_{2}}{1+1/\eta_{2}}.
\end{equation}
Assuming that $\rho_{\text{h},1}$ is known we can take $\eta_1$ from
this equation and substitute it into the second equation in \eqref{GGE-TBA-0}.
This way we obtain the following closed set of equations:
\begin{equation}
\begin{split}
\label{GGE-TBA}
    \log\eta_2&=d\star \left[
\log\left(
\frac{d+d\star \frac{\rho_{2}}{1+1/\eta_{2}}}{d+d\star \frac{\rho_{2}}{1+1/\eta_{2}}-\rho_{1,\text{h}}}
\right)+\log(1+\eta_{3})
\right]
\\
  \rho_2&=d\star\left[ \rho_{1,\text{h}}+\frac{\rho_{3}}{1+1/\eta_{3}}
\right]\\
  \log\eta_j&=d\star \left[
\log(1+\eta_{j-1})+\log(1+\eta_{j+1})
\right],\qquad j>2\\
 \rho_j&=d\star\left[ \frac{\rho_{j-1}}{1+1/\eta_{j-1}}+\frac{\rho_{j+1}}{1+1/\eta_{j+1}}
\right], \qquad j>2.\\
\end{split}
\end{equation}
These equations can be solved numerically by simultaneous iteration
which we found to converge very well (in particular no damping of
the iterations was required).  Once the solution is found the
correlators can be computed using the method described in Appendix \ref{sec:appcorr}.

It is useful to consider the asymptotic behaviour of the solution of
the GGE-TBA, which is the same as in the purely thermal case. For large $j$ the $\eta_j(u)$ become
approximately constant functions. Therefore we obtain the algebraic equations
\begin{equation}
\label{ezazami}
  \eta_j^2=(1+\eta_{j-1})(1+\eta_{j+1}).
\end{equation}
The general solution at zero magnetization is \cite{takahashi-book}
\begin{equation}
\label{etaasymp}
  \eta_j=(j+a)^2-1,
\end{equation}
where $a$ is an arbitrary real number. At infinite temperature
(or equivalently if all $\beta_j$ of the GGE vanish)  equation
\eqref{ezazami} is valid at arbitrary $j$ and we obtain $a=1$.
On the other hand, at zero temperature ($\beta_1=\infty$) we obtain
the ground state which consists of 1-strings only with no holes and
therefore $a=0$. In the general case $a$ depends on all of the
Lagrange multipliers.

For higher strings the density functions also become rapidity
independent and the total densities
$\rho_{j}=\rho_{\text{r},j}+\rho_{\text{h},j}$ satisfy
\begin{equation}
\rho_j  =\frac{1}{2} \left(\rho_{j-1} \frac{(j-1+a)^2-1}{(j-1+a)^2}+
\rho_{j+1} \frac{(j+1+a)^2-1}{(j+1+a)^2}
\right).
\end{equation}
The physically relevant, decaying solution to this equation is
\begin{equation}
\label{rhoasymp}
  \rho_j=\alpha \frac{j+a}{(j+a)^2-1},
\end{equation}
where $\alpha$ depends on the particular situation. With the help of equations
\eqref{etaasymp} and \eqref{rhoasymp} a very efficient iteration
scheme can be constructed which produces high accuracy numerical
results such that only a small number of equations is kept from the
infinite system \eqref{GGE-TBA}. Details of this method are described
in Appendix \ref{sec:iter}.

In the dimer case the generating function for the charges was calculated in
\cite{fagotti-collura-essler-calabrese}. In our normalizations it reads
\begin{equation}
  G_{D}=-\sinh(\eta)
\frac{4\cos(2\lambda)(\sinh^2(\eta)-\cosh(\eta))+\cosh(\eta)+2\cosh(2\eta)+3\cosh(3\eta)-2}
{4(\cosh(2\eta)-\cos(2\lambda))^2}.
\end{equation}
Inverting the convolution \eqref{rho1hG} we obtained
\begin{equation}
\label{rho1hD}
  \rho_{1,\text{hole}}+ s_1=
-\frac{D}{2}( s_1+ s_3)+\frac{C}{2}( \tilde s_1+ \tilde s_3),
\end{equation}
where
\begin{equation}
  \begin{split}
  s_3(u)&=-\frac{\sinh(3\eta)}{\sin(u+3i\eta/2)\sin(u-3i\eta/2)}
  \\
\tilde s_1(u)&= -\frac{1}{2}\left(
\frac{1}{\sin^2(u+i\eta/2)}+\frac{1}{\sin^2(u-i\eta/2)}\right)\\
\tilde s_3(u)&= -\frac{1}{2}\left(
\frac{1}{\sin^2(u+3i\eta/2)}+\frac{1}{\sin^2(u-3i\eta/2)}
\right)
  \end{split}
\end{equation}
and
\begin{equation}
  \begin{split}
    D&=
-\frac{\sinh(\eta)}{4\sinh^3(2\eta)}\left(
\frac{3}{2}\cosh(5\eta)+\cosh(4\eta)+\frac{1}{2}\cosh(3\eta)-2\cosh(\eta)-1
\right)
\\
C&=
-\frac{\sinh(\eta)}{4\sinh^2(2\eta)}\left(
\cosh(4\eta)+\cosh(3\eta)-\cosh(\eta)-1
\right)
  \end{split}
\end{equation}
We substituted \eqref{rho1hD} into \eqref{GGE-TBA} and computed
numerical results for short range correlators for various values of
$\Delta$. Examples of the results are given in Table \ref{tab:1}. Our
numerical results agree with those obtained in
\cite{fagotti-collura-essler-calabrese} and they differ from the QA
predictions \cite{sajat-oTBA}.
\begin{table}
  \centering
  \begin{tabular}{|c||c|c|c|}
\hline
$\Delta$ & $\vev{\sigma_1^z\sigma_2^z}$ & $\vev{\sigma_1^z\sigma_3^z}$
& $\vev{\sigma_1^z\sigma_4^z}$ \\
\hline
1.4 & -0.5583393723 &   0.2531046021 & -0.1427967954 \\
\hline
1.6 & -0.5750941519 &   0.2793140206 &  -0.1574652764 \\
\hline
2 &  -0.5918972433 &   0.3079627101 &  -0.1652651609  \\
\hline
3 & -0.5943245488 &   0.3170656703 &  -0.1405690303 \\
\hline
4 &  -0.5841620734 &   0.3044963585 &  -0.1117786185 \\
\hline
5 &  -0.5739043321 &   0.2908962124 &  -0.0900289420 \\
\hline
  \end{tabular}
\caption{The GGE predictions for short range $z-z$ correlators for
the quench starting from the dimer state.}
\label{tab:1}
\end{table}

It is interesting to consider the Quench Action evaluated at the GGE
saddle point, which gives information about its spectral weight in
\eqref{funcint} and therefore its relevance for the quench 
dynamics\footnote{The idea to evaluate the QA for the GGE solution was
  suggested to us by M\'arton Mesty\'an.}. If a non-zero value of the QA
is found then this alone proves that the GGE can not describe the
asymptotic steady state of the system. Quite surprisingly an even stronger
result holds: In this particular quench problem the QA is infinite for
any configuration where the root densities have the thermal
asymptotics for higher strings, and this includes the GGE
solution. Our statement follows
follows from the fact that for large $j$ we have 
\begin{equation}
\label{rhorootasymp}
  \rho_{\text{r,j}}=\frac{\rho_{\text{r,j}}+\rho_{\text{h,j}}}{1+\eta_j}
\sim \frac{1}{j^3},
\end{equation}
and the large $j$ asymptotics of the overlap source is \cite{sajat-oTBA}
\begin{equation*}
  g_j\approx -\eta j^2.
\end{equation*}
It is easy to see that the entropy is finite,
therefore the overlap terms make expression \eqref{QA} infinite if
the asymptotics \eqref{rhorootasymp} 
 holds. The physical meaning of this finding is that the spectral
 weight of these states in \eqref{DE} decays faster than exponentially as a function
 of the volume. We note that the same behaviour is found for the
 quantum quench starting from the N\'eel state.

\subsection{Generating new configurations}

\label{sec:further}

The discrepancy between the predictions of the QA and GGE-TBA methods
is already a sign of the failure of the GETH. Both methods give the
same set of local charges, and therefore they share the hole density for the
1-strings. However, the other hole densities $\rho_{\text{h},j}$,
$j>2$ are different and therefore the correlation functions are
different. This phenomenon was already observed in \cite{JS-oTBA} for
the quantum quench starting from the N\'eel state.

Here we show that an arbitrary large family of configurations can be
generated which share the mean values of the charges but can have
arbitrary correlation functions. The easiest way to construct new
states is by altering the hole densities themselves. For example we
can start with the QA solution $\{\rho^{QA}_{\text{h},j}\}$ and modify a
handful of the hole densities, calculate the new root densities from
\eqref{rhoelso} and the correlators from the formulas in Appendix
\ref{sec:appcorr}. Alternatively one could interpolate between the QA
and GGE solutions as
\begin{equation*}
  \rho_{\text{h},j}(\beta)=(1-\beta)\rho^{QA}_{\text{h},j}+\beta\rho^{GGE}_{\text{h},j},\quad
\text{where}\quad \beta=0\dots 1,
\end{equation*}
such that all correlations would interpolate between their QA and GGE
values. 

\begin{figure}
  \centering
  \begin{subfigure}[b]{8cm}
    \includegraphics[width=\textwidth]{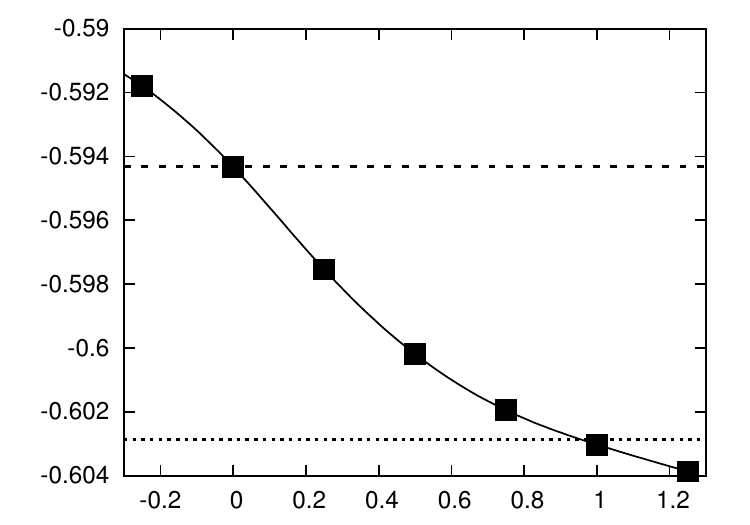}
                \caption{$\vev{\sigma_1^z\sigma_2^z}$}
                \label{fig:zz2}
  \end{subfigure}
  \begin{subfigure}[b]{8cm}
    \includegraphics[width=\textwidth]{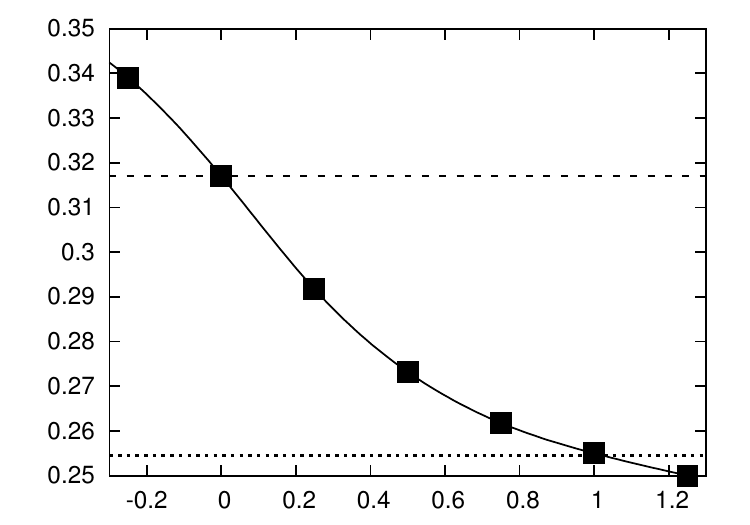}
                \caption{$\vev{\sigma_1^z\sigma_3^z}$}
                \label{fig:zz3}
  \end{subfigure}
   \caption{Examples for short range correlators for $\Delta=3$ as a function of
     $\beta_2$, which is the coupling of the artificial source term in
     \eqref{haj2}. The data points correspond to different root
     configurations which have the same conserved charges as the dimer
     initial state, but have different local correlations.  The dashed
     and the dotted lines correspond to the 
     GGE and QA predictions, respectively. The former were obtained
     from the GGE-TBA system \eqref{GGE-TBA}, whereas the latter were taken from
     \cite{sajat-oTBA}. The point $\beta_2=0$ is
     corresponds to the GGE solution, whereas at $\beta_2=1$ the
     correlators are close, but not identical to the QA
     prediction. The numerical data refer directly to the
     thermodynamic limit, therefore they directly demonstrate the failure of
     the GETH.}\label{fig:animals}
\end{figure}

For our numerical examples we choose a similar but numerically more
convenient method. We consider an artificial modification of the
GGE-TBA system \eqref{GGE-TBA} such that we add new source terms to
a finite number of the equations. The sources can be arbitrary
functions, but we chose the functions which appear in the overlap-TBA \eqref{oTBA}.
This way we obtain
configurations which share the thermal asymptotics but whose
correlation functions can be arbitrarily close to those of the QA saddle
point. For the numerical calculations we considered the equations
\begin{equation}
\begin{split}
\label{haj2}
    \log\eta_2&=\beta_2 f_2 +d\star \left[
\log\left(
\frac{d+d\star \frac{\rho_{2}}{1+1/\eta_{2}}}{d+d\star \frac{\rho_{2}}{1+1/\eta_{2}}-\rho_{1,\text{h}}}
\right)+\log(1+\eta_{3})
\right]
\\
  \rho_2&=d\star\left[ \rho_{1,\text{h}}+\frac{\rho_{3}}{1+1/\eta_{3}}
\right]\\
  \log\eta_j&=d\star \left[
\log(1+\eta_{j-1})+\log(1+\eta_{j+1})
\right],\qquad j>2\\
 \rho_j&=d\star\left[ \frac{\rho_{j-1}}{1+1/\eta_{j-1}}+\frac{\rho_{j+1}}{1+1/\eta_{j+1}}
\right], \qquad j>2.\\
\end{split}
\end{equation}
Here a new source term is added only to the equation for $\eta_2$. The
function $f_2$ is the source of the overlap-TBA defined in \eqref{fj},
and $\beta_2$ is an arbitrary real number. At $\beta_2=0$ we obtain
the GGE solution, whereas for $\beta_2=1$ the
resulting densities (for small $j$) are more similar to the solution of the
overlap-TBA. Note that gradually adding all higher source terms $f_j$ a
point-wise convergence to the overlap-TBA solution could be
achieved. Surprisingly, already the addition of $f_2$ results in
correlators which are similar to the exact QA predictions. Numerical
results as a function of $\beta_2$ are shown in Figure \ref{fig:animals}.

\section{Conclusions}

\label{sec:conclusions}

In this work we demonstrated that the Generalized Eigenstate
Thermalization Hypothesis is not valid  in the XXZ
spin chain. Following \cite{andrei-gge} it can be argued that the GETH
breaks down in all models with multiple particle species. As a result,
the GGE does not give correct predictions for the stationary values of
local correlators. This follows from the fact that the squared
overlaps which are the statistical weights in the
diagonal ensemble are unrelated to the generalized Boltzmann 
weights of the GGE, and if the GETH fails, the states selected by the
two ensembles typically yield different correlation functions. In
fact, in the absence of the GETH it would be an exception to find
examples where the GGE would predict the correct stationary state. It is
important to stress that all our results refer directly to the infinite volume
limit and therefore there are no finite size effects to be considered.

On a technical level we argued that in the XXZ spin chain the hole
densities of the strings should be regarded as fundamental variables,
from which all other data (including the root densities and correlation
functions) can be derived. It was already shown in
\cite{JS-oTBA} that the mean values of local charges only depend on the
hole density of the 1-strings, and here we argued that the correlators depend on all
higher $\rho_{\text{h,j}}$. Therefore, a large family of new root
configurations with different correlations but the same local charges
can be produced by artificially changing the higher hole densities in
an arbitrary way. We presented an example for this procedure, with the
numerical results shown in Fig. \ref{fig:animals}.

In Subsection \ref{sec:GGE-TBA} we calculated the GGE predictions for
the quench from the dimer  state using the TBA method. Our numerical
data agree with those obtained in
\cite{fagotti-collura-essler-calabrese} using the QTM method. This
agreement supports the validity of both approaches. 

As an interesting by-product of our calculations we found that the
Quench Action evaluated at the GGE saddle point solution is 
infinite. As a result, the spectral weight of the GGE solution (and in
fact of all states with thermal asymptotics) decays faster than
exponential as a function of the volume. In simple terms this means
that the GGE solution is very far from the states selected by the
Quench Action, which actually determine the dynamics of the system.

Throughout this work the expression GGE was used in the conventional
sense, meaning that the generalized statistical ensemble includes the
local charges which are obtained by taking derivatives of the usual
transfer matrix. New quasi-local operators have been found recently for
the regime $\Delta<1$ \cite{Prosen1,Prosen2,kvazilok2} and it is
interesting open question whether the addition of these new operators
to the GGE is enough to fix all hole densities to their correct
values, thus making the GGE complete and correct. Unfortunately the
construction of \cite{Prosen1,Prosen2,kvazilok2} does not produce
local operators for $\Delta>1$, and it remains to be seen whether new
charges can be produced by other means in this regime.

\vspace{1cm}
{\bf Acknowledgements} 

\bigskip

We are grateful to G\'abor Tak\'acs and M\'arton Kormos for many
useful discussions, which inspired the present work. Also, we are
thankful to M\'arton Mesty\'an for his suggestion to evaluate the
Quench Action at the GGE-TBA saddle point solution. We are also
grateful to Jean-S\'ebastien Caux and Jacopo De Nardis for useful
discussions and for sending examples of their numerical data to us. We
would like to thank M\'arton Kormos for valuable suggestions on the manuscript.

\bigskip

\appendix

\section{Correlation functions}

\label{sec:appcorr}

In this section we show how to compute local correlators in the XXZ
spin chain using the functions $\rho_{\text{h},k}(\lambda)$, which are
the densities of holes for the $k$-strings. As a first step one
computes the root densities $\rho_{\text{h},k}(\lambda)$ from
\eqref{rhoelso} and the functions 
$\eta_k(\lambda)=\rho_{\text{h},k}(\lambda)/\rho_{\text{r},k}(\lambda)$. The
$\eta_k$ are then used as the input to further linear equations.

Let us define
\begin{equation}
\begin{split}
d^{(a)}(u)=\left(\frac{\partial }{\partial u}\right)^{a} d(u)\qquad\qquad
\tilde d^{(a)}(u)=\left(\frac{\partial }{\partial u}\right)^{a}
\tilde d(u),
\end{split}
\end{equation}
where $d(u)$ is given by \eqref{du} and
\begin{equation}
\begin{split}
\tilde d(u)=
-\sum_{n=1}^\infty \sin(2n u)   \frac{\sinh(\eta n)}{\cosh^2(\eta n)}.
\end{split}
\end{equation}
These functions satisfy the relations
\begin{equation}
  \frac{\partial }{\partial \eta} d^{(a)}(u)=
\frac{\partial }{\partial u} \tilde d^{(a)}(u).
\end{equation}
Let us introduce auxiliary functions $\rho^{(a)}$ and $\sigma^{(a)}$
which satisfy
\begin{equation}
\label{rhosigma}
\begin{split}
\rho^{(a)}_k(u)&= \delta_{k,1}d^{(a)}(u)+
d\star \left(
\frac{\rho^{(a)}_{k-1}}{1+1/\eta_{k-1}}+
\frac{\rho^{(a)}_{k+1}}{1+1/\eta_{k+1}}
\right)\\
\sigma^{(a)}_k(u)&=- \delta_{k,1}\tilde d^{(a)}(u)-
\tilde d \star 
\left(
\frac{\rho^{(a)}_{k-1}}{1+1/\eta_{k-1}}+
\frac{\rho^{(a)}_{k+1}}{1+1/\eta_{k+1}}
\right)+\\
&\hspace{3cm}+ d \star \left(
\frac{\sigma^{(a)}_{k-1}}{1+1/\eta_{k-1}}+
\frac{\sigma^{(a)}_{k+1}}{1+1/\eta_{k+1}}
\right).
\end{split}
\end{equation}
Note that the equation for $\rho^{(0)}_k$ coincides with
\eqref{rhoelso}, therefore it can be identified as
 $\rho^{(0)}_k=\rho_{\text{r,k}}+\rho_{\text{h,k}}$.

Using the notation
\begin{equation}
  f\cdot g =\int_{-\pi/2}^{\pi/2} \frac{du}{2\pi} f(u) g(u)
\end{equation}
we define the quantities
\begin{equation}
\begin{split}
\label{OmGa}
 \Omega_{a,b}&= 
-2 (-1)^{(a+b)/2}\left((-1)^a G_{a+b}+
d^{(b)} \cdot \frac{\rho^{(a)}_1}{1+1/\eta_1}
\right)\\
  \Gamma_{a,b}&=
2(-1)^{(a+b-1)/2}\left((-1)^b\tilde G_{a+b}
+\tilde d^{(b)} \cdot \frac{\rho^{(a)}_1}{1+1/\eta_1}
+d^{(b)}_1 \cdot \frac{\sigma^{(a)}_1}{1+1/\eta_1}\right),
\end{split}
\end{equation}
where
\begin{equation}
G_{a}=
  -2\sum_{n=-\infty}^\infty  \frac{(2ni)^{a}}{1+e^{2\eta |n|}}
\qquad\qquad
\tilde G_{a}=-\sum_{n=-\infty}^\infty  \frac{|n|(2ni)^{(a-1)}}{\cosh^2(n\eta)}.
\end{equation}
The following symmetry properties hold:
\begin{equation*}
   \Omega_{a,b}= \Omega_{b,a}\qquad\qquad
 \Gamma_{a,b}=- \Gamma_{b,a}.
\end{equation*}
Furthermore, $\Omega_{a,b}$ is non-vanishing if $a+b=0 \mod 2$, whereas
$\Gamma_{a,b}$ is non-vanishing if $a+b=1 \mod 2$.

It can be shown that the auxiliary functions and the quantities
$\Omega_{a,b}$ and $\Gamma_{a,b}$ are identical to those defined in
\cite{sajat-corr}. The advantage of the above formulas over those of
\cite{sajat-corr} is that here only neighbouring equations are coupled
and $\Omega_{a,b}$ and $\Gamma_{a,b}$ are expressed using a single
integral. This makes the numerical evaluation much more effective.

The numbers $\omega_{a,b}$ and $W_{a,b}$ are computed using
\begin{equation}
\label{omWW}
  \begin{split}
    \omega_{a,b}&=-(-1)^{(a+b)/2}\Omega_{a,b} -(-1)^b\frac{1}{2}
    \left(\frac{\partial}{\partial u}\right)^{a+b} K(u)\Big|_{u=0}\\
W_{a,b}&=
-(-1)^{(a+b-1)/2}\Gamma_{a,b} +(-1)^b\frac{1}{2}
    \left(\frac{\partial}{\partial u}\right)^{a+b} \tilde K(u)\Big|_{u=0},
  \end{split}
\end{equation}
where
\begin{equation*}
  K(u)=\frac{\sinh2\eta }{\sinh(u+\eta)\sinh(u-\eta)}\qquad\qquad
 \tilde  K(u)=\frac{\sinh(2u) }{\sinh(u+\eta)\sinh(u-\eta)}.
\end{equation*}
Finally, local correlators are obtained by substituting $\omega_{a,b}$
and $W_{a,b}$ into the already available formulas of the QTM
literature. Examples for such formulas can be found in
\cite{XXZ-massive-corr-numerics-Goehmann-Kluemper}. The results for
the nearest neighbour and next-to-nearest neighbour z-z correlators
are
\begin{equation}
\label{corrpeldak}
  \begin{split}
\vev{\sigma^z_1\sigma^z_2}_T&=\coth(\eta)\omega_{0,0}+W_{1,0}\\
\vev{\sigma^z_1\sigma^z_3}_T&=2\coth(2\eta)\omega_{0,0}+W_{1,0}+\tanh(\eta)\frac{\omega_{2,0}-2\omega_{1,1}}{4}
-\frac{\sinh^2(\eta)}{4}W_{2,1}.
  \end{split}
\end{equation}

It follows from the original definitions in \cite{sajat-corr} that the numbers $\Omega_{a,0}$ are
proportional to the local charges $\vev{Q_a}$. Constructing the generating function
$G(\lambda)$ defined in \eqref{G} the formula \eqref{rho1hG} is easily
derived from \eqref{OmGa}.

\section{Numerical solution of the TBA equations with thermal asymptotics}

\label{sec:iter}

Here we describe a simple method to for the truncation of the infinite
system \eqref{GGE-TBA} which produces very accurate numerical data
with a low number of equations. The basic idea is to use the exact
asymptotics given by \eqref{etaasymp} and \eqref{rhoasymp} by fixing
the parameters $a$ and $\alpha$ at each iteration from one of the
lower equations and to substitute them into the last equation. 

If the infinite set is truncated to $N_t$ equations, then 
the iterations are performed in the usual way for $j=1\dots N_t-1$
 and for the last equation we use
\begin{equation}
 \log \eta_{N_t}=d\star \log(1+\eta_{N_t-1})+\log(N_t+a+1), 
\end{equation}
where $a$ is extracted from the average of one of the other pseudo-energies:
\begin{equation}
\label{aaa}
a=    \sqrt{\left(\int \frac{du}{\pi}\eta_{N_t-b}(u)\right)+1} +b-N_t.
\end{equation}
We can choose for example $b=2$. Concerning the iterations for $\rho_j$ we use
\begin{equation}
\label{ite2}
  \rho_{N_t}=d\star\frac{\rho_{N_t-1}}{1+1/\eta_{N_t-1}}
+ 
\frac{\alpha}{2(N_t+1+a)},
\end{equation}
where the parameter $\alpha$ can be fixed from one of the earlier
equations similar to \eqref{aaa}. The linear equations for the
auxiliary functions $\rho^{(a)}_k$ and $\sigma^{(a)}_k$ needed for the
correlation functions (see Appendix \ref{sec:appcorr}) can be iterated
using \eqref{ite2} as well.

We observed that with this method correlation functions can be
obtained up to a precision of at least $10^{-11}$ with a small number
of equations. For example at $\Delta=2$ it was enough to choose
$N_t=10$, whereas for $\Delta=1.4$ we chose $N_t=14$. 
It was checked in all cases that the results for the correlators
do not change as we vary $N_t$ or $b$, or the resolution of the
rapidity axis for the integrations.

\bigskip

\addcontentsline{toc}{section}{References}
\bibliography{../../pozsi-general}

\providecommand{\href}[2]{#2}\begingroup\raggedright\begin{thebibliography}{10}

\bibitem{Silva-quench-colloquium}
A.~Polkovnikov, K.~Sengupta, A.~Silva, and M.~Vengalattore,
  ``\textit{Colloquium} : Nonequilibrium dynamics of closed interacting quantum
  systems,'' \href{http://dx.doi.org/10.1103/RevModPhys.83.863}{{\em Rev. Mod.
  Phys.} {\bf 83} (2011)  863--883}.

\bibitem{huse-review}
R.~{Nandkishore} and D.~A. {Huse}, ``{Many body localization and thermalization
  in quantum statistical mechanics},'' {\em ArXiv e-prints} (2014)  ,
  \href{http://arxiv.org/abs/1404.0686}{{\tt arXiv:1404.0686
  [cond-mat.stat-mech]}}.

\bibitem{cold-atom-review}
I.~Bloch, J.~Dalibard, and S.~Nascimb{\`e}ne, ``Quantum simulations with
  ultracold quantum gases,'' \href{http://dx.doi.org/10.1038/nphys2259}{{\em
  Nat. Phys.} {\bf 8} (2012)  267}.

\bibitem{eth1}
J.~M. Deutsch, ``Quantum statistical mechanics in a closed system,''
  \href{http://dx.doi.org/10.1103/PhysRevA.43.2046}{{\em Physical Review A}
  {\bf 43} (1991) no.~4, 2046--2049}.

\bibitem{eth2}
M.~{Srednicki}, ``{Chaos and quantum thermalization},''
  \href{http://dx.doi.org/10.1103/PhysRevE.50.888}{{\em Physical Review E} {\bf
  50} (1994)  888--901},
  \href{http://arxiv.org/abs/arXiv:cond-mat/9403051}{{\tt
  arXiv:cond-mat/9403051}}.

\bibitem{rigol-eth}
M.~{Rigol}, V.~{Dunjko}, and M.~{Olshanii}, ``{Thermalization and its mechanism
  for generic isolated quantum systems},''
  \href{http://dx.doi.org/10.1038/nature06838}{{\em Nature} {\bf 452} (2008)
  854--858}, \href{http://arxiv.org/abs/0708.1324}{{\tt arXiv:0708.1324
  [cond-mat.stat-mech]}}.

\bibitem{rigol-gge}
M.~Rigol, V.~Dunjko, V.~Yurovsky, and M.~Olshanii, ``Relaxation in a Completely
  Integrable Many-Body Quantum System: An Ab Initio Study of the Dynamics of
  the Highly Excited States of 1D Lattice Hard-Core Bosons,''
  \href{http://dx.doi.org/10.1103/PhysRevLett.98.050405}{{\em Physical Review
  Letters} {\bf 98} (2007) no.~5, 050405},
  \href{http://arxiv.org/abs/arXiv:cond-mat/0604476}{{\tt
  arXiv:cond-mat/0604476}}.

\bibitem{rigol-GETH}
A.~C. Cassidy, C.~W. Clark, and M.~Rigol, ``Generalized Thermalization in an
  Integrable Lattice System,''
  \href{http://dx.doi.org/10.1103/PhysRevLett.106.140405}{{\em Physical Review
  Letters} {\bf 106} (2011)  140405},
  \href{http://arxiv.org/abs/1008.4794}{{\tt arXiv:1008.4794
  [cond-mat.stat-mech]}}.

\bibitem{free-gge-1}
T.~{Barthel} and U.~{Schollw{\"o}ck}, ``{Dephasing and the Steady State in
  Quantum Many-Particle Systems},''
  \href{http://dx.doi.org/10.1103/PhysRevLett.100.100601}{{\em Physical Review
  Letters} {\bf 100} (2008) no.~10, 100601},
  \href{http://arxiv.org/abs/0711.4896}{{\tt arXiv:0711.4896
  [cond-mat.stat-mech]}}.

\bibitem{free-gge-2}
M.~{Kollar} and M.~{Eckstein}, ``{Relaxation of a one-dimensional Mott
  insulator after an interaction quench},''
  \href{http://dx.doi.org/10.1103/PhysRevA.78.013626}{{\em Physical Review A}
  {\bf 78} (2008) no.~1, 013626}, \href{http://arxiv.org/abs/0804.2254}{{\tt
  arXiv:0804.2254 [cond-mat.str-el]}}.

\bibitem{free-gge-3}
M.~A. {Cazalilla}, A.~{Iucci}, and M.-C. {Chung}, ``{Thermalization and quantum
  correlations in exactly solvable models},''
  \href{http://dx.doi.org/10.1103/PhysRevE.85.011133}{{\em Physical Review E}
  {\bf 85} (2012) no.~1, 011133}, \href{http://arxiv.org/abs/1106.5206}{{\tt
  arXiv:1106.5206 [cond-mat.stat-mech]}}.

\bibitem{free-gge-4}
A.~{Iucci} and M.~A. {Cazalilla}, ``{Quantum quench dynamics of the Luttinger
  model},'' \href{http://dx.doi.org/10.1103/PhysRevA.80.063619}{{\em Physical
  Review A} {\bf 80} (2009) no.~6, 063619},
  \href{http://arxiv.org/abs/0903.1205}{{\tt arXiv:0903.1205
  [cond-mat.str-el]}}.

\bibitem{ising-quench-1}
P.~{Calabrese}, F.~H.~L. {Essler}, and M.~{Fagotti}, ``{Quantum quench in the
  transverse field Ising chain: I. Time evolution of order parameter
  correlators},''
  \href{http://dx.doi.org/10.1088/1742-5468/2012/07/P07016}{{\em Journal of
  Statistical Mechanics: Theory and Experiment} {\bf 7} (2012)  16},
  \href{http://arxiv.org/abs/1204.3911}{{\tt arXiv:1204.3911
  [cond-mat.quant-gas]}}.

\bibitem{ising-quench-2}
P.~{Calabrese}, F.~H.~L. {Essler}, and M.~{Fagotti}, ``{Quantum quenches in the
  transverse field Ising chain: II. Stationary state properties},''
  \href{http://dx.doi.org/10.1088/1742-5468/2012/07/P07022}{{\em Journal of
  Statistical Mechanics: Theory and Experiment} {\bf 7} (2012)  22},
  \href{http://arxiv.org/abs/1205.2211}{{\tt arXiv:1205.2211
  [cond-mat.stat-mech]}}.

\bibitem{essler-truncated-gge}
M.~{Fagotti} and F.~H.~L. {Essler}, ``{Reduced Density Matrix after a Quantum
  Quench},'' \href{http://dx.doi.org/10.1103/PhysRevB.87.245107}{{\em Physical
  Review B} {\bf 87} (2013) no.~24, 245107},
  \href{http://arxiv.org/abs/1302.6944}{{\tt arXiv:1302.6944
  [cond-mat.stat-mech]}}.

\bibitem{JS-oTBA}
B.~{Wouters}, M.~{Brockmann}, J.~{De Nardis}, D.~{Fioretto}, and J.-S. {Caux},
  ``{From N{\'e}el to XXZ: exact solution from the quench action},'' {\em ArXiv
  e-prints} (2014)  , \href{http://arxiv.org/abs/1405.0172}{{\tt
  arXiv:1405.0172 [cond-mat.str-el]}}.

\bibitem{quench-action}
J.-S. {Caux} and F.~H.~L. {Essler}, ``{Time Evolution of Local Observables
  After Quenching to an Integrable Model},''
  \href{http://dx.doi.org/10.1103/PhysRevLett.110.257203}{{\em Physical Review
  Letters} {\bf 110} (2013) no.~25, 257203},
  \href{http://arxiv.org/abs/1301.3806}{{\tt arXiv:1301.3806
  [cond-mat.stat-mech]}}.

\bibitem{sajat-oTBA}
B.~{Pozsgay}, M.~{Mesty{\'a}n}, M.~A. {Werner}, M.~{Kormos}, G.~{Zar{\'a}nd},
  and G.~{Tak{\'a}cs}, ``{Correlations after quantum quenches in the XXZ spin
  chain: Failure of the Generalized Gibbs Ensemble},'' {\em ArXiv e-prints}
  (2014)  , \href{http://arxiv.org/abs/1405.2843}{{\tt arXiv:1405.2843
  [cond-mat.stat-mech]}}.

\bibitem{fagotti-collura-essler-calabrese}
M.~Fagotti, M.~Collura, F.~H.~L. Essler, and P.~Calabrese, ``Relaxation after
  quantum quenches in the spin-$\frac{1}{2}$ Heisenberg XXZ chain,''
  \href{http://dx.doi.org/10.1103/PhysRevB.89.125101}{{\em Physical Review B}
  {\bf 89} (2014)  125101}.

\bibitem{andrei-gge}
G.~{Goldstein} and N.~{Andrei}, ``{Failure of the GGE hypothesis for integrable
  models with bound states},'' {\em ArXiv e-prints} (2014)  ,
  \href{http://arxiv.org/abs/1405.4224}{{\tt arXiv:1405.4224
  [cond-mat.quant-gas]}}.

\bibitem{korepin-book}
V.~Korepin, N.~Bogoliubov, and A.~Izergin, {\em Quantum inverse scattering
  method and correlation functions}.
\newblock Cambridge University Press, 1993.

\bibitem{XXX}
H.~Bethe, ``Zur Theorie der Metalle,'' {\em Zeitschrift {f\"ur} Physik} {\bf
  A71} (1931)  205.

\bibitem{LM-sajat}
B.~{Pozsgay}, ``{Mean values of local operators in highly excited Bethe
  states},'' \href{http://dx.doi.org/10.1088/1742-5468/2011/01/P01011}{{\em J.
  Stat. Mech.} {\bf 2011} (2011)  P01011},
  \href{http://arxiv.org/abs/1009.4662}{{\tt arXiv:1009.4662 [hep-th]}}.

\bibitem{sajat-corr}
M.~{Mesty{\'a}n} and B.~{Pozsgay}, ``{Short distance correlators in the XXZ
  spin chain for arbitrary string distributions},'' {\em ArXiv e-prints} (2014)
   , \href{http://arxiv.org/abs/1405.0232}{{\tt arXiv:1405.0232
  [cond-mat.stat-mech]}}.

\bibitem{takahashi-book}
M.~Takahashi, {\em Thermodynamics of One-Dimensional Solvable Models}.
\newblock Cambridge University Press, 1999.

\bibitem{essler-xxz-gge}
M.~{Fagotti} and F.~H.~L. {Essler}, ``{Stationary behaviour of observables
  after a quantum quench in the spin-1/2 Heisenberg XXZ chain},''
  \href{http://dx.doi.org/10.1088/1742-5468/2013/07/P07012}{{\em Journal of
  Statistical Mechanics: Theory and Experiment} {\bf 7} (2013)  12},
  \href{http://arxiv.org/abs/1305.0468}{{\tt arXiv:1305.0468
  [cond-mat.stat-mech]}}.

\bibitem{XXZ-finite-T-factorization}
H.~E. Boos, F.~G\"ohmann, A.~Kl\"{u}mper, and J.~Suzuki, ``Factorization of the
  finite temperature correlation functions of the XXZ chain in a magnetic
  field,'' \href{http://dx.doi.org/10.1088/1751-8113/40/35/001}{{\em J. Phys.
  A} {\bf 40} (2007)  10699}, \href{http://arxiv.org/abs/0705.2716}{{\tt
  arXiv:0705.2716}}.

\bibitem{XXZ-factorization-recent-osszefoglalo}
J.~{Sato}, B.~{Aufgebauer}, H.~{Boos}, F.~{G{\"o}hmann}, A.~{Kl{\"u}mper},
  M.~{Takahashi}, and C.~{Trippe}, ``{Computation of Static Heisenberg-Chain
  Correlators: Control over Length and Temperature Dependence},''
  \href{http://dx.doi.org/10.1103/PhysRevLett.106.257201}{{\em Physical Review
  Letters} {\bf 106} (2011) no.~25, 257201},
  \href{http://arxiv.org/abs/1105.4447}{{\tt arXiv:1105.4447
  [cond-mat.str-el]}}.

\bibitem{MG}
C.~K. Majumdar and D.~K. Ghosh, ``On Next-Nearest-Neighbor Interaction in
  Linear Chain. I,'' \href{http://dx.doi.org/10.1063/1.1664978}{{\em Journal of
  Mathematical Physics} {\bf 10} (1969) no.~8, 1388--1398}.

\bibitem{sajat-xxz-gge}
B.~Pozsgay, ``The generalized Gibbs ensemble for Heisenberg spin chains,'' {\em
  Journal of Statistical Mechanics: Theory and Experiment} {\bf 2013} (2013)
  no.~07, 3, \href{http://arxiv.org/abs/1304.5374}{{\tt arXiv:1304.5374
  [cond-mat.stat-mech]}}.

\bibitem{Caux-Neel-overlap2}
M.~{Brockmann}, J.~{De Nardis}, B.~{Wouters}, and J.-S. {Caux}, ``{N{\'e}el-XXZ
  state overlaps: odd particle numbers and Lieb-Liniger scaling limit},'' {\em
  ArXiv e-prints} (2014)  , \href{http://arxiv.org/abs/1403.7469}{{\tt
  arXiv:1403.7469 [cond-mat.stat-mech]}}.

\bibitem{caux-stb-LL-BEC-quench}
J.~De~Nardis, B.~Wouters, M.~Brockmann, and J.-S. Caux, ``Solution for an
  interaction quench in the Lieb-Liniger Bose gas,''
  \href{http://dx.doi.org/10.1103/PhysRevA.89.033601}{{\em Physical Review A}
  {\bf 89} (2014)  033601}, \href{http://arxiv.org/abs/1308.4310}{{\tt
  arXiv:1308.4310 [cond-mat.stat-mech]}}.

\bibitem{Caux-Neel-overlap1}
M.~{Brockmann}, J.~{De Nardis}, B.~{Wouters}, and J.-S. {Caux}, ``{A
  Gaudin-like determinant for overlaps of N{\'e}el and XXZ Bethe states},''
  \href{http://dx.doi.org/10.1088/1751-8113/47/14/145003}{{\em Journal of
  Physics A Mathematical General} {\bf 47} (2014) no.~14, 145003},
  \href{http://arxiv.org/abs/1401.2877}{{\tt arXiv:1401.2877
  [cond-mat.stat-mech]}}.

\bibitem{sajat-neel}
B.~{Pozsgay}, ``{Overlaps between eigenstates of the XXZ spin-1/2 chain and a
  class of simple product states},''
  \href{http://dx.doi.org/10.1088/1742-5468/2014/06/P06011}{{\em J. Stat.
  Mech.} {\bf 2014} (2013) no.~6, P06011},
  \href{http://arxiv.org/abs/1309.4593}{{\tt arXiv:1309.4593
  [cond-mat.stat-mech]}}.

\bibitem{JS-private}
J.-S. Caux and J.~{De Nardis}, ``private communication,''.

\bibitem{Prosen1}
T.~{Prosen}, ``{Open XXZ Spin Chain: Nonequilibrium Steady State and a Strict
  Bound on Ballistic Transport},''
  \href{http://dx.doi.org/10.1103/PhysRevLett.106.217206}{{\em Physical Review
  Letters} {\bf 106} (2011) no.~21, 217206},
  \href{http://arxiv.org/abs/1103.1350}{{\tt arXiv:1103.1350
  [cond-mat.str-el]}}.

\bibitem{Prosen2}
T.~{Prosen}, ``{Quasilocal conservation laws in XXZ spin-1/2 chains: open,
  periodic and twisted boundary conditions},'' {\em ArXiv e-prints} (2014)  ,
  \href{http://arxiv.org/abs/1406.2258}{{\tt arXiv:1406.2258 [math-ph]}}.

\bibitem{kvazilok2}
R.~G. {Pereira}, V.~{Pasquier}, J.~{Sirker}, and I.~{Affleck}, ``{Exactly
  conserved quasilocal operators for the XXZ spin chain},'' {\em ArXiv
  e-prints} (2014)  , \href{http://arxiv.org/abs/1406.2306}{{\tt
  arXiv:1406.2306 [cond-mat.stat-mech]}}.

\bibitem{XXZ-massive-corr-numerics-Goehmann-Kluemper}
C.~{Trippe}, F.~{G{\"o}hmann}, and A.~{Kl{\"u}mper}, ``{Short-distance thermal
  correlations in the massive XXZ chain},''
  \href{http://dx.doi.org/10.1140/epjb/e2009-00417-7}{{\em European Physical
  Journal B} {\bf 73} (2010)  253--264},
  \href{http://arxiv.org/abs/0908.2232}{{\tt arXiv:0908.2232
  [cond-mat.str-el]}}.

\end{thebibliography}\endgroup
\bibliographystyle{utphys}

\end{document}